\documentclass[onecolumn]{revtex4}
\usepackage{amsmath,amssymb,graphics,epsfig,subfigure}
\usepackage{color}
\usepackage[colorlinks,linkcolor=red,anchorcolor=red,citecolor=green]{hyperref}
\usepackage{setspace}
\usepackage{booktabs}
\usepackage{float}
\begin{document}

\thispagestyle{empty}

\begin{center}

\title{Fokker-Planck equation for black holes in thermal potential}

\author{Zhen-Ming Xu\footnote{E-mail: xuzhenm@nwu.edu.cn}
        \vspace{6pt}\\}

\affiliation{ $^{1}$Institute of Modern Physics, Northwest University, Xi'an 710127, China\\
$^{2}$School of Physics, Northwest University, Xi'an 710127, China\\
$^{3}$Shaanxi Key Laboratory for Theoretical Physics Frontiers, Xi'an 710127, China\\
$^{4}$Peng Huanwu Center for Fundamental Theory, Xi'an 710127, China}

\begin{abstract}
We construct a kind of thermal potential and then put the black hole thermodynamic system in it. In this regard, some thermodynamic properties of the black hole are related to the geometric characteristics of the thermal potential. Driven by the intrinsic thermodynamic fluctuations, the behavior of the black hole in the thermal potential is stochastic. With the help of solving the Fokker-Planck equation analytically, we obtain the discrete energy spectrum of Schwarzschild and Banados–Teitelboim–Zanelli (BTZ) black holes in the thermal potential. For Schwarzschild black hole, the energy spectrum is proportional to the temperature of the ensemble, which is an external parameter, and the ground state is non-zero. For BTZ black hole, the energy spectrum only depends on the AdS radius, which is the intrinsic parameter. Moreover, the ground state of BTZ black hole in thermal potential is zero. This also reflects the difference between three-dimensional gravity and four-dimensional gravity.
\end{abstract}

\maketitle
\end{center}

\section{Introduction}
As an important bridge between general relativity and quantum mechanics, black hole has attracted great attention in its related physical properties. The intriguing discovery of the semi-classical description on black hole temperature and entropy~\cite{Bekenstein1973,Bardeen1973} revealed the thermal nature of black holes and established a profound relationship between gravity and thermodynamics. Thus, black hole is mapped to a thermodynamic system. Thermodynamics or statistical physics may provide a potentially and complementary description of black hole physics. Even it can provide some original insights into the quantum nature of gravity. From then on, some abundant properties of black holes have been investigated~\cite{Kastor2009,Dolan2011,Kubiznak2017}.

Based on the remarkable observation that the horizon area of non-extremal black holes behaves as a classical adiabatic invariant, Bekenstein conjectured that the horizon area of a non-extremal quantum black hole should have a discrete eigenvalue spectrum~\cite{Bekenstein1974}. After that, within the discussion of the quasinormal modes and Bohr’s correspondence principle, some works~\cite{Hod1998,Maggiore2008} have shown that the mass and horizon area of black holes have a discrete spectrum. Afterwards, many studies suggests that black holes exhibit some kind of quantum behavior and make the relationship between the black hole and the quantization of gravity closer~\cite{Banerjee2010a,Majhi2010,Banerjee2010b,Majhi2011,Bakshi2017}.

Nowadays, with the development of research in this respect, black hole, an unique thermodynamic system, presents incredible and peculiar thermodynamic properties with some quantum characteristics, such as Hawking-Page phase transition~\cite{Hawking1983} corresponding to confinement-deconfinement transition in gauge theory~\cite{Witten1998} with the AdS/CFT correspondence, large and small black hole phase transition similar to gas-liquid phase transition~\cite{Kubiznak2012,Bhattacharya2020,Dehyadegari2019,Bhattacharya2017a,Bhattacharya2017b,Majhi2017}, information loss of black hole~\cite{Raju2020}, chaotic effect~\cite{Maldacena2016,Dalui2019,Dalui2020a,Dalui2020b,Dalui2020c,Majhi2021,Dalui2021}, and etc. Some phenomenological schemes have been proposed to analyze the thermodynamic properties of black holes. Black hole molecular hypothesis~\cite{Wei2015} based on thermodynamics geometry~\cite{Ruppeiner2014} is used to analyze some possible micro-behaviors of black holes~\cite{Miao2018,Wei2019,Guo2019,Xu2020a,Wei2020,Xu2020b}. With the idea of stochastic process~\cite{Zwanzig2001}, the dynamic process of black hole phase transition have been discussed~\cite{Li2020,Li2020b,Wei2021,Cai2021}, where the on-shell Gibbs free energy is generalized to the off-shell Gibbs free energy by
replacing the Hawking temperature with the ensemble temperature. In addition, in our previous work~\cite{Xu2021}, we introduced the general Landau potential to analyze the process of black hole phase transition dynamically.

In this paper, our main motivations are as follows.
\begin{itemize}
  \item Based on the spirit of free energy landscape~\cite{Li2020}, we want to find a more intuitive way to analyze some dynamic behaviors of black holes. Therefore, we construct a kind of thermal potential by using the temperature of the ensemble. We put the black hole system in such unique thermal potential. Some geometric characteristics of the thermal potential can directly reflect thermodynamic properties of the black hole. Driven by thermal fluctuations, black hole system moves in such a thermal potential. Moreover, we find that the thermal potential constructed in current way is equivalent in form to the off-shell free energy in the free energy landscape, but both of them have some certain different physical connotations.
  \item We are also interested in the dynamical scenario of the black hole in thermal potential, but we are more concerned about whether we can get an analytic result. Fortunately, in the process of the stochastic effect of thermal fluctuation, we obtain the analytic dynamic description of Schwarzschild and BTZ black holes by solving the Fokker-Planck equation analytically.
  \item Our results also suggest an essential difference between three-dimensional gravity and four-dimensional gravity. The four-dimensional Schwarzschild black hole is in the inverted harmonic oscillator potential, so its thermal stability is self-evident. The eigenvalue spectrum of Schwarzschild black hole in inverted harmonic oscillator potential is proportional to the temperature of the ensemble. At the same time, the system has a non-zero ground state. Once the ensemble temperature is exactly in accordance with the expression of Schwarzschild black hole temperature, i.e., the thermal potential reaches the maximum, the eigenvalue is proportional to the reciprocal of the mass of Schwarzschild black hole. While for three-dimensional BTZ black hole, it is in harmonic oscillator potential and its energy spectrum only depends on the AdS radius. Furthermore, the ground state of BTZ black hole in harmonic oscillator potential is zero.
\end{itemize}

This paper is organized as follows. In Section~\ref{sec2}, we construct the thermal potential for black holes by introducing the canonical ensemble. In Section~\ref{sec3}, we give a brief introduction of the Fokker-Planck equation and its transformation to eigenvalue problem. Taking two specific black holes (Schwarzschild and BTZ black holes) as examples, we solve the Fokker-Planck equations of these two black holes analytically, and give the corresponding energy spectrum and related discussions. At last, Section~\ref{sec4} is devoted to summary and some future prospects. Throughout this paper, we adopt the units $\hbar=c=k_{_{B}}=G=1$.

\section{Thermal potential}\label{sec2}
Consider a canonical ensemble at temperature $T$ composed of a large number of states in which one or a group of them can represent a real black hole. The real black hole state (on-shell) is the solution of the Einstein field equation while others (off-shell) are not. When the ensemble temperature $T$ is equal to the Hawking temperature $T_h$, the ensemble is made up of real black hole states, which is in equilibrium. For a specific black hole thermodynamic system, we can construct the thermal potential
\begin{eqnarray}\label{tpotential}
U=\int (T_h-T)\text{d}S,
\end{eqnarray}
where the thermodynamic entropy $S$ of the black hole is seen as a variable. For black holes, we know $T_h=t(S,Y)$, where the function $t(S,Y)$ is the relation satisfied by thermodynamic entropy $S$ and other parameters $Y$ of the black hole, like the AdS radius $l$, charge $Q$, angular momentum $J$, etc. The ensemble temperature $T$ now here is treated as an independent constant, which can take any positive value in any way. Note that $T=T_h$ mentioned in this paper is just one of the ways to get the value of ensemble temperature $T$.

The integrand in the above definition~(\ref{tpotential}) of thermal potential can be understood as the deviation of all possible states in the canonical ensemble from the real black hole state (or the equilibrium state). In other words, in the equilibrium state, the thermal potential will show extreme behavior, i.e.,
\begin{eqnarray}
\frac{\text{d}U}{\text{d}S}=0 \quad \Rightarrow \quad T=T_h.
\end{eqnarray}

Physically, through the construction of such a thermal potential~(\ref{tpotential}), we put a black hole thermodynamic system in the potential field $U$. Due to the thermodynamic fluctuations, the stochastic behavior of black holes in such a potential field can reflect some thermodynamic characteristics of black holes.

Moreover, another advantage of the thermal potential~(\ref{tpotential}) is that the concavity and convexity at the extreme point are related to the stability of the thermodynamic system,
\begin{eqnarray}
\delta\left(\left.\frac{\text{d}U}{\text{d}S}\right|_{T=T_h}\right)=\left.\frac{\partial t(S,Y)}{\partial S}\right|_Y\delta S.
\end{eqnarray}
When $\partial t(S,Y)/\partial S>0$, the thermodynamic system is in a stable state, while $\partial t(S,Y)/\partial S <0$ corresponds to an unstable state.

For a simple thermodynamic system, according to the first law of thermodynamics $dE=T_h dS-PdV$, where $E$ is the internal energy, $P$ is the pressure, and $V$ is the thermodynamic volume of the system, we have
\begin{eqnarray}
U=\int (T_h-T)\text{d}S=E+PV-TS.
\end{eqnarray}
Formally, we can see that the thermal potential constructed in this paper is equivalent to the off-shell free energy in the free energy landscape~\cite{Li2020}.

\section{Fokker-Planck equation}\label{sec3}

Due to thermal fluctuations, black hole behaves in the thermal potential, which leads to different phase transition characteristics. In fact, this is a kind of stochastic process. The probability
distribution $W(x,t)$ of these black hole states (including on shell states and off-shell states) evolving in time under the thermal fluctuation should be described by the probabilistic Fokker-Planck equation (or in the mathematical literature, it is also called a forward Kolmogorov equation). The Fokker-Planck equation provides a powerful tool with which the effects of fluctuations close transition points can be adequately treated. It is not restricted to systems near thermal equilibrium, as well can be applied to systems far from thermal equilibrium (like the laser).

One-variable Fokker-Planck equation with time-independent drift coefficient $D^{(1)}(x)$ and constant diffusion coefficient $D$ is~\cite{Zwanzig2001,Risken1988}
\begin{eqnarray}\label{fp}
\frac{\partial W(x,t)}{\partial t}=\left[\frac{\partial}{\partial x}f^{'}(x)+D\frac{\partial^2}{\partial x^2}\right]W(x,t)=L_{_{\text{FP}}}W(x,t)=-\frac{\partial}{\partial x} S(x,t),
\end{eqnarray}
where $S(x,t)$ is the probability current, the potential $f(x)=-\int^{x}D^{(1)}(y)dy$ and $f^{'}(x):=df(x)/dx$. A separation ansatz for probability density $W(x,t)=\varphi(x)e^{-\varepsilon t}$ leads to the eigenvalue equation for the Fokker-Planck equation with appropriate boundary conditions
\begin{eqnarray}\label{efp}
L_{_{\text{FP}}}\varphi(x)=-\varepsilon\varphi(x).
\end{eqnarray}

For convenience, we introduce $\Phi(x)=f(x)/D$ resulting that the Fokker-Planck operator $L_{_{\text{FP}}}$ can be written as
\begin{eqnarray}
L_{_{\text{FP}}}=\frac{\partial}{\partial x}De^{-\Phi(x)}\frac{\partial}{\partial x}e^{\Phi(x)}.
\end{eqnarray}
Easily, we can obtain an Hermitian operator $L:=-e^{\Phi(x)/2}L_{_{\text{FP}}}e^{-\Phi(x)/2}$ and the eigenvalue equation~(\ref{efp}) becomes~\cite{Risken1988}
\begin{eqnarray}\label{efp2}
L\psi(x)=\varepsilon\psi(x),
\end{eqnarray}
where $\psi(x)=e^{\Phi(x)/2}\varphi(x)$ and the Hermitian operator $L$ has the same form as the single-particle Hamilton operator in quantum mechanics,
\begin{eqnarray}
L=-D\frac{\partial^2}{\partial x^2}+V_s(x), \qquad V_s(x)=\frac{1}{4D}\left[f^{'}(x)\right]^2-\frac12 f^{''}(x).
\end{eqnarray}

Now we talk about the boundary conditions for the above eigenvalue problem of the Fokker-Planck equation. We note that we study in this paper the motion behavior of all possible states in the canonical ensemble in the thermal potential and the thermal potential is constructed from the black hole background. Once the thermal potential is determined, this becomes an usually quantum mechanical problem. Hence the below boundary conditions are natural. For all possible states in the canonical ensemble, the real black hole states are only some special cases of them.
\begin{itemize}
  \item Reflecting boundary condition (RBC): in the region $x>x_{\text{max}}$ or $x<x_{\text{min}}$, the potential $\Phi(x)$ tends to an infinite high positive value, which requires $S=0$.
  \item Absorbing boundary condition (ABC): in the region $x>x_{\text{max}}$ or $x<x_{\text{min}}$, the potential $\Phi(x)$ tends to an infinite large negative value, which requires $e^{\Phi}W =0$.
  \item Natural boundary condition (NBC): for $x_{\text{max}}\rightarrow +\infty$ and $x_{\text{min}} \rightarrow -\infty$, we have $S=0$ or $e^{\Phi}W =0$.
\end{itemize}
In the next discussion, we will see the stochastic behaviors of two different black holes (four-dimensional Schwarzschild black hole and three-dimensional BTZ black hole) in different thermal potentials. By solving the Fokker-Planck equation, we will see the application of the above boundary conditions.

\subsection{Application 1: Schwarzschild black hole}
For four-dimensional Schwarzschild black hole, its metric reads~\cite{Kastor2009}
\begin{eqnarray}
\text{d}s^2=-\left(1-\frac{2M}{r}\right)\text{d}t^2+\frac{\text{d}r^2}{1-2M/r}+r^2 (\text{d}\theta^2+\sin^2 \theta \text{d}\varphi^2),
\end{eqnarray}
where $M$ is the ADM mass of the black hole. Correspondingly, the Hawking temperature and  Bekenstein-Hawking entropy of Schwarzschild black hole take the forms (in equilibrium)~\cite{Kastor2009} in terms of the radius of the event horizon $r_h$
\begin{eqnarray}\label{tands}
T_h=\frac{1}{4\pi r_h}, \qquad S=\pi r_h^2.
\end{eqnarray}

In the light of Eq.~(\ref{tpotential}), we can obtain the thermal potential of the Schwarzschild black hole easily
\begin{eqnarray}\label{schp}
U=\frac{1}{2}r_h-\pi T r_h^2.
\end{eqnarray}

It is the inverted harmonic oscillator potential or parabolic potential barrier~\footnote{For Eq.~(\ref{schp}), we can rewrite it as
      \begin{eqnarray*}
      U=-\pi T \tilde{r_h}^2+\frac{1}{16\pi T}, \qquad \tilde{r_h}=r_h-\frac{1}{4\pi T}.
      \end{eqnarray*}
      Hence we can called it as inverted harmonic oscillator (IHO) potential once the constant term $\frac{1}{16\pi T}$ in the above formula is absorbed within $U$.}, which plays indispensable roles in physics particularly in unstable systems. If the ensemble temperature $T$ is exactly in accordance with the temperature expression~(\ref{tands}) of Schwarzschild black hole, i.e., $T=T_h$, we can clearly see that the thermal potential reaches the maximum.

For Schwarzschild black hole, it is now in the potential field~(\ref{schp}). Next, in order to analyze some dynamical behaviors of the black hole, here we set the potential $f(x)=x/2-\pi T x^2$, and thus the effective potential is
\begin{eqnarray}
V_s(x)=\frac{\pi^2 T^2}{D}z^2+\pi T, \qquad z=x-\frac{1}{4\pi T}.
\end{eqnarray}

The Fokker-Planck equation~(\ref{efp2}) becomes the following simple form with the help of auxiliary variable $\xi=\sqrt{\pi T/D}z$
\begin{eqnarray}
\frac{\partial^2}{\partial \xi^2}\psi(x)+\left(\frac{\varepsilon}{\pi T}-1-\xi^2\right)\psi(x)=0.
\end{eqnarray}

Hence we can get the corresponding eigenvalues and eigenfunctions
\begin{eqnarray}
\varepsilon_n=2\pi T(n+1), \qquad n=0,1,2,\cdots
\end{eqnarray}
\begin{eqnarray}
\psi_n(x)=\left(\frac{T}{D}\right)^{1/4}\frac{1}{\sqrt{2^n n!}}H_n(\xi)e^{-\xi^2/2},
\end{eqnarray}
where $H_n(\xi)$ are the Hermite polynomials. Now, we make some discussions about above results.
\begin{itemize}
  \item We find that when the Schwarzschild black hole in the potential field, due to thermal fluctuations, its eigenvalues are discrete, and are proportional to the ensemble temperature under the ABC and NBC at $x\rightarrow x_{\text{max}}, x_{\text{min}}\rightarrow \pm\infty$. The higher the ensemble temperature is, the greater its eigenvalue is.
  \item The system has non-zero ground state, which is the characteristic of the inverted harmonic oscillator potential. The ground state of Schwarzschild black hole in thermal potential is
  \begin{eqnarray}
  \varepsilon_0=2\pi T.
  \end{eqnarray}
  When the ensemble temperature is higher, the corresponding ground state is larger.
  \item The difference between the two energy levels of the system is also proportional to the ensemble temperature,
  \begin{eqnarray}
  \varepsilon_{n+1}-\varepsilon_n=2\pi T.
  \end{eqnarray}
  That is to say, the difference between two adjacent energy levels is always the same as the ground state energy of the system.
  \item If the inverted harmonic oscillator described by~(\ref{schp}) in one-dimensional quantum mechanics, the Lyapunov exponent in both classical and quantum mechanics is $\lambda=\sqrt{2\pi T}$~\cite{Maldacena2016,Hashimoto2020,Morita2021}. Therefore, the eigenvalue of Schwarzschild black hole in the thermal potential is proportional to the square of the Lyapunov exponent of the inverse harmonic oscillator~\footnote{We need to note that here we investigate the motion behavior of all possible states in the canonical ensemble in the thermal potential, while the real black hole states are only some special cases of all states in the canonical ensemble, i.e., the cases of the ensemble temperature $T$ being equal to the Hawking temperature $T_h$. In this framework, we get the Lyapunov exponent $\lambda=\sqrt{2\pi T}$ where $T$ is the ensemble temperature (an external parameter).},
   \begin{eqnarray}
  \varepsilon_n \propto \lambda^2,
  \end{eqnarray}
   which is a very meaningful result and helpful for us to understand some chaotic effects of the black hole system.
  \item We have known that for the inverted harmonic oscillator potential, no stationary exists for the Fokker-Planck equation. However, when we consider ABC and NBC at $x\rightarrow x_{\text{max}}, x_{\text{min}}\rightarrow \pm\infty$, eigenfunctions can do exist (the probability current $S$ for these eigenfunctions is finite), and they can be used to calculate the transition probability. Immediately, we obtain the transition probability into eigenfunctions $(t\geq t^{'})$
\begin{eqnarray}
P\left(z,t|z^{'},t^{'}\right)=\sqrt{\frac{T}{D[1-e^{-4\pi T(t-t^{'})}]}}\exp\left(-\frac{\pi T[z-e^{-2\pi T(t-t^{'})}z^{'}]^2}{D[1-e^{-4\pi T(t-t^{'})}]}\right)e^{-2\pi T(t-t^{'})}.
\end{eqnarray}
  \item If the ensemble temperature is exactly in accordance with the temperature expression of Schwarzschild black hole, that is $T=T_h=1/(8\pi M)$, we can find that the eigenvalue of Schwarzschild black hole in the thermal potential is proportional to the inverse of the mass
\begin{eqnarray}\label{mas}
\varepsilon_n=\frac{n+1}{4M}, \qquad n=0,1,2,\cdots,
\end{eqnarray}
indicating that the larger the mass of a black hole is, the smaller its eigenvalues is in the thermal potential.
\end{itemize}

\subsection{Application 2: BTZ black hole}
Next, we take a look at a three-dimensional black hole. As a simple example, we consider a neutral and non-rotating Banados–Teitelboim–Zanelli (BTZ) black hole. Its metric is~\cite{Banados1992,Frassino2015}
\begin{eqnarray}
\text{d}s^2 &=&-\left(-2m+\frac{r^2}{l^2}\right)\text{d}t^2+\frac{\text{d}r^2}{-2m+r^2/l^2}+r^2 \text{d}\varphi^2,
\end{eqnarray}
where $m$ is related to the black hole mass, $l$ is the AdS radius which is connected with the negative cosmological constant $\Lambda$ via $\Lambda=-1/l^2$. Naturally, some basic thermodynamic properties of the BTZ black hole in terms of the event horizon radius $r_h$ are
\begin{eqnarray}
T_h=\frac{r_h}{2\pi l^2}, \qquad S=\frac12 \pi r_h.
\end{eqnarray}

With the help of Eq.~(\ref{tpotential}), we can obtain the thermal potential of the BTZ black hole
\begin{eqnarray}\label{btzp}
U=\frac{r_h^2}{8l^2}-\frac{\pi T}{2}r_h.
\end{eqnarray}
Obviously, BTZ black hole is in harmonic oscillator potential~\footnote{For Eq.~(\ref{btzp}), we can rewrite it as
      \begin{eqnarray*}
      U=\frac{\tilde{r_h}^2}{8l^2}-\frac{\pi^2 T^2}{2l^2}, \qquad \tilde{r_h}=r_h-2\pi T l^2.
      \end{eqnarray*}
      Hence we can called it as harmonic oscillator potential once the constant term $-\frac{\pi^2 T^2}{2l^2}$ in the above formula is absorbed within $U$.} and it is always thermodynamically stable. Formally, we set $f(x)=x^2/(8l^2)-\pi T x/2$, and then the effective potential is
\begin{eqnarray}
V_s(x)=\frac{z^2}{64Dl^4}-\frac{1}{8l^2}, \qquad z=x-2\pi T l^2.
\end{eqnarray}

By substituting auxiliary variable $\xi=\sqrt{1/(8Dl^2)}z$, the thermodynamic evolution equation~(\ref{efp2}) of BTZ black hole can be written in the following simple form
\begin{eqnarray}
\frac{\partial^2}{\partial \xi^2}\psi(x)+(8l^2\varepsilon+1-\xi^2)\psi(x)=0.
\end{eqnarray}
Hence we can get the corresponding eigenvalues and eigenfunctions about the thermodynamic evolution of BTZ black hole in the thermal potential
\begin{eqnarray}
\varepsilon_n=\frac{n}{4l^2}, \qquad n=0,1,2,\cdots
\end{eqnarray}
\begin{eqnarray}
\psi_n(x)=\left(\frac{1}{8\pi D l^2}\right)^{1/4}\frac{1}{\sqrt{2^n n!}}H_n(\xi)e^{-\xi^2/2}.
\end{eqnarray}

\begin{itemize}
  \item When we get the above eigenvalues, we use RBC and NBC at $x\rightarrow x_{\text{max}}, x_{\text{min}}\rightarrow \pm\infty$. Because the BTZ black hole is in the harmonic oscillator potential, we get the stationary solution.
  \item The energy spectrum of BTZ black hole only depends on the parameter of the black hole itself, the AdS radius $l$. This is different from the case of Schwarzschild black hole, where the energy spectrum is related to the temperature of the ensemble. Furthermore, the energy spectrum is also discrete.
  \item Although BTZ black hole is in harmonic oscillator potential, its behavior is different from that of quantum harmonic oscillator. The most prominent one is that the ground state is zero, i.e., $\varepsilon_0=0$.
  \item The difference between two adjacent energy levels is always $1/(4l^2)$.
\end{itemize}

\section{Summary}\label{sec4}
We consider a canonical ensemble composed of a large number of states with the same structure under the same macro condition, i.e., at the same temperature $T$.  One or a group of states can represent the real black hole systems and its temperature can be labeled as the Hawking temperature $T_h$. The real black hole states (on-shell states) are solutions of the Einstein field equation, while others (off-shell states) are not. When the ensemble temperature $T$ is equal to the Hawking temperature $T_h$, the ensemble is made up of real black hole states, which is in equilibrium. When the ensemble temperature $T$ is not equal to the Hawking temperature $T_h$, we can say that all possible states in the canonical ensemble deviate from the real black hole state (or the equilibrium state). Note that the ensemble temperature $T$ can take any positive value in any way and $T=T_h$ is just one of the ways to get the value of the ensemble temperature $T$.

In usual thermodynamics, in free energy, the degree of the thermal motion is measured by the product of temperature and the entropy. Therefore, we simply introduce a thermal potential~(\ref{tpotential}) to roughly reflect the degree of deviation. In the equilibrium state, the thermal potential shows extreme behavior. Meanwhile the concavity and convexity of potential can be related to the stability of thermodynamic system. In this way, we can observe the thermal motion behavior of states of the canonical ensemble in such a thermal potential due to the thermal fluctuation.

With the help of solving the Fokker-Planck equation analytically in such thermal potential, we investigate the motion behavior of all possible states in the canonical ensemble in the thermal potential. This becomes the barrier penetration problem (for thermal potential of Schwarzschild black hole) in quantum mechanics or the motion of states in the potential well (for thermal potential of BTZ black hole). Therefore, it is natural to calculate the eigenvalues and transition probability of states in the potential barrier or potential well.

For Schwarzschild black hole, it is in inverted harmonic oscillator potential. We consider the absorbing and natural boundary conditions to obtain eigenvalues. The calculation shows that the motion behavior of the states in the canonical ensemble we consider in the Schwarzschild thermal potential Eq.~(\ref{schp}) is discrete, and its eigenvalue depends on the temperature $T$ of the canonical ensemble. The higher the ensemble temperature is, the larger the eigenvalue is. At the same time, the system has a non-zero ground state. When the ensemble temperature $T$ is exactly in accordance with the temperature $T_h$ of Schwarzschild black hole, i.e., states in the canonical ensemble are at the highest point of the barrier (or the thermal potential Eq.~(\ref{schp}) reaches the maximum), the ensemble is made up of real black hole states and the eigenvalue of Schwarzschild black hole state in such thermal potential is proportional to the inverse of the mass, i.e. Eq.~(\ref{mas}). That is to say, the larger the black hole mass is, the smaller the eigenvalue is and the smaller the energy value of the ground state is.

For BTZ black hole, it is in harmonic oscillator potential. We consider the reflecting and natural boundary conditions to obtain eigenvalues. Compared with Schwarzschild black hole and quantum harmonic oscillator, BTZ black hole presents two major differences. One is that the energy spectrum only depends on the intrinsic parameter, the AdS radius, of the black hole. Other is that the ground state is zero. This further reflects the unique nature of three-dimensional gravity.

Naturally, the current method can be extended to other black hole models, such as the thermal potential of several simple black holes listed in the table~\ref{tab}. Although the expression of thermal potential of other models is somewhat complex, we can use perturbation method to calculate the eigenvalues and eigenfunctions of the system, and then obtain some quantum properties of thermodynamic system due to thermal fluctuation. In addition, we believe that the current way can also be extended to various statistical physical models to obtain many properties of the system.

\begin{table}[!htbp]
\centering
\begin{tabular}{c|c}
\hline
  &Thermal potential $f(x)=$\\
\hline
\rule{0pt}{22pt} Schwarzschild-AdS black hole&  $\dfrac{1}{2}x+\dfrac{4\pi P}{3}x^3-\pi T x^2$  \\
\rule{0pt}{22pt} Reissner-Nordstr\"{o}m black hole &$\dfrac{1}{2}x+\dfrac{Q^2}{2x}-\pi T x^2$   \\
\rule{0pt}{22pt} Charged AdS black hole &$\dfrac{1}{2}x+\dfrac{4\pi P}{3}x^3+\dfrac{Q^2}{2x}-\pi T x^2$   \\
\rule{0pt}{22pt} Charged BTZ black hole &$\dfrac{1}{8l^2}x^2-\dfrac{\pi T}{2}x-\dfrac{Q^2}{16}\ln x$   \\
\rule{0pt}{22pt} Rotating BTZ black hole &$\dfrac{1}{8l^2}x^2-\dfrac{\pi T}{2}x+\dfrac{J^2}{32x^2}$   \\
\\
\hline
\end{tabular}
\caption{The expressions of thermal potential of several simple black holes.}
\label{tab}
\end{table}

\section*{Acknowledgments}
ZMX would like to thank Professor Shao-Wen Wei for his meaningful discussion. Project funded by National Natural Science Foundation of China (Grant Nos. 12105222, 12047502 and 11947208), China Postdoctoral Science Foundation (Grant No. 2020M673460). This research is supported by The Double First-class University Construction Project of Northwest University. The author would like to thank the anonymous referee for the helpful comments that improve this work greatly.


\begin{thebibliography}{99}

\bibitem{Bekenstein1973}J.D. Bekenstein, Black holes and entropy, Phys. Rev. D 7, 2333 (1973).

\bibitem{Bardeen1973}J.M. Bardeen, B. Carter, and S. Hawking, The four laws of black hole mechanics, Commun. Math. Phys. 31, 161 (1973).

\bibitem{Kastor2009}D. Kastor, S. Ray, and J. Traschen, Enthalpy and the mechanics of AdS black holes, Class. Quant. Grav. 26, 195011 (2009).

\bibitem{Dolan2011}B. P. Dolan, The cosmological constant and black-hole thermodynamic potentials, Class. Quant. Grav. 28, 125020 (2011).

\bibitem{Kubiznak2017}D. Kubiznak, R.B. Mann, and M. Teo, Black hole chemistry: thermodynamics with Lambda, Class. Quant. Grav. 34, 063001 (2017).

\bibitem{Bekenstein1974}J.D. Bekenstein, The Quantum Mass Spectrum of the Kerr Black Hole,  Lett. Nuovo Cimento 11, 467 (1974).

\bibitem{Hod1998}S. Hod, Bohr’s Correspondence Principle and the Area Spectrum of Quantum Black Holes, Phys. Rev. Lett. 81, 4293 (1998).

\bibitem{Maggiore2008}M. Maggiore, Physical Interpretation of the Spectrum of Black Hole Quasinormal Modes, Phys. Rev. Lett. 100, 141301 (2008).

\bibitem{Banerjee2010a}R. Banerjee, B.R. Majhi, E.C. Vagenas, Quantum tunneling and black hole spectroscopy, Phys. Lett. B 686, 279 (2010).

\bibitem{Majhi2010}B.R. Majhi, Hawking radiation and black hole spectroscopy in Horava-Lifshitz gravity, Phys. Lett. B 686, 49 (2010).

\bibitem{Banerjee2010b}R. Banerjee, C. Kiefer, B.R. Majhi, Quantum gravitational correction to the Hawking temperature from the Lemaitre-Tolman-Bondi model, Phys. Rev. D 82, 044013 (2010).

\bibitem{Majhi2011}B.R. Majhi, E.C. Vagenas, Black hole spectroscopy via adiabatic invariance, Phys. Lett. B 701, 623 (2011).

\bibitem{Bakshi2017}A. Bakshi, B.R. Majhi, S. Samanta, Gravitational surface Hamiltonian and entropy quantization, Phys. Lett.  B 765, 334 (2017) .

\bibitem{Hawking1983}S. Hawking and D.N. Page, Thermodynamics of black holes in anti-de Sitter space, Commun. Math. Phys. 87, 577 (1983).

\bibitem{Witten1998}E. Witten, Anti-de Sitter space, thermal phase transition, and confinement in gauge theories, Adv. Theor. Math. Phys. 2, 505 (1998).

\bibitem{Kubiznak2012}D. Kubiznak and R.B. Mann, $P-V$ criticality of charged AdS black holes, JHEP 07, 033 (2012).

\bibitem{Bhattacharya2020}K. Bhattacharya, B.R. Majhi, Thermogeometric study of van der Waals like phase transition in black holes: an alternative approach, Phys. Lett. B 802,  135224 (2020).

\bibitem{Dehyadegari2019}A. Dehyadegari, B.R. Majhi, A. Sheykhi, A. Montakhab, Universality class of alternative phase space and Van der Waals criticality, Phys. Lett. B 791, 30 (2019).

\bibitem{Bhattacharya2017a}K. Bhattacharya, B.R. Majhi, S. Samanta, Van der Waals criticality in AdS black holes: a phenomenological study, Phys. Rev. D 96, 084037 (2017).

\bibitem{Bhattacharya2017b}K. Bhattacharya, B.R. Majhi, Thermogeometric description of the van der Waals like phase transition in AdS black holes, Phys. Rev. D 95, 104024 (2017).

\bibitem{Majhi2017}B.R. Majhi, S. Samanta, P-V criticality of AdS black holes in a general framework, Phys. Lett. B 773,  203 (2017).

\bibitem{Raju2020}S. Raju, Lessons from the Information Paradox, arXiv:2012.05770 [hep-th].

\bibitem{Maldacena2016}J. Maldacena, S.H. Shenker and D. Stanford, A bound on chaos, JHEP 08, 106 (2016).

\bibitem{Dalui2019}S. Dalui, B.R. Majhi, P. Mishra, Presence of horizon makes particle motion chaotic, Phys. Lett. B 788,  486 (2019).

\bibitem{Dalui2020a}S. Dalui, B.R. Majhi, P. Mishra, Induction of chaotic fluctuations in particle dynamics in a uniformly accelerated frame, Int. J. Mod. Phys. A 35,  2050081 (2020).

\bibitem{Dalui2020b}S. Dalui, B.R. Majhi, P. Mishra, Horizon induces instability locally and creates quantum thermality, Phys. Rev. D 102, 044006 (2020).

\bibitem{Dalui2020c}S. Dalui, B.R. Majhi, Near horizon local instability and quantum thermality, Phys. Rev. D 102, 124047 (2020).

\bibitem{Majhi2021}B.R. Majhi, Is randomness near a black hole key for thermalization of its horizon? arXiv: 2101.04458[gr-qc].

\bibitem{Dalui2021}S. Dalui, B.R. Majhi, Horizon thermalization of Kerr black hole through local instability, arXiv: 2103.11613 [gr-qc].

\bibitem{Wei2015}S.-W. Wei and Y.-X. Liu, Insight into the microscopic structure of an AdS black hole from a thermodynamical phase transition, Phys. Rev. Lett. 115, 111302 (2015); Erratum ibid. 116, 169903 (2016).

\bibitem{Ruppeiner2014}G. Ruppeiner, Thermodynamic curvature and black holes, in Breaking of Supersymmetry and Ultraviolet Divergences in Extended Supergravity, edited by S. Bellucci, Springer Proceedings in Physics Vol. 153, Springer, New York (2014).

\bibitem{Miao2018}Y.-G. Miao and Z.-M. Xu, Thermal molecular potential among micromolecules in charged AdS black holes, Phys. Rev. D 98, 044001 (2018).

\bibitem{Wei2019}S.-W. Wei, Y.-X. Liu, and R.B. Mann, Repulsive interactions and universal properties of charged anti-de Sitter black hole microstructures, Phys. Rev. Lett. 123, 071103 (2019).

\bibitem{Guo2019}X.-Y. Guo, H.-F. Li, L.-C. Zhang and R. Zhao, Microstructure and continuous phase transition of a Reissner-NordstromAdS black hole, Phys. Rev. D 100, 064036 (2019).

\bibitem{Xu2020a}Z.-M. Xu, B. Wu, and W.-L. Yang, Ruppeiner thermodynamic geometry for the Schwarzschild AdS black hole, Phys. Rev. D 101, 024018 (2020).

\bibitem{Wei2020}S.-W. Wei, Y.-X. Liu, and R.B. Mann, Novel dual relation and constant in Hawking-Page phase transition, Phys. Rev. D 102, 104011 (2020).

\bibitem{Xu2020b}Z.-M. Xu, The correspondence between thermodynamic curvature and isoperimetric theorem from ultraspinning black hole, Phys. Lett. B 807, 135529 (2020).

\bibitem{Zwanzig2001}R. Zwanzig, Nonequilibrium atatistical mechanics, Oxford University Press (2001).

\bibitem{Li2020}R. Li and J. Wang, Thermodynamics and kinetics of Hawking-Page phase transition, Phys. Rev. D 102, 024085 (2020).

\bibitem{Li2020b}R. Li, K. Zhang and J. Wang, Thermal dynamic phase transition of Reissner-Nordström Anti-de Sitter black holes on free energy landscape, JHEP 10, 090 (2020).

\bibitem{Wei2021}S.-W. Wei, Y.-Q. Wang, Y.-X. Liu, and R.B. Mann, Observing dynamic oscillatory behavior of triple points among black hole thermodynamic phase transitions, Sci. China-Phys. Mech. Astron. 64, 270411 (2021).

\bibitem{Cai2021}R.-G. Cai, Oscillatory behaviors near a black hole triple point, Sci. China-Phys. Mech. Astron. 64, 290432 (2021).

\bibitem{Xu2021}Z.-M. Xu, B. Wu, and W.-L. Yang, van der Waals fluid and charged AdS black hole in the Landau theory, Class. Quant. Grav. 38, 205008 (2021).

\bibitem{Risken1988}H. Risken, The Fokker-Planck equation: methods of solution and applications, Second Edition, Springer (1988).

\bibitem{Hashimoto2020}K. Hashimoto, K.-B. Huh, K.-Y. Kim, and R. Watanabe, Exponential growth of out-of-time-order correlator without chaos: inverted harmonic oscillator, JHEP 11, 068 (2020).

\bibitem{Morita2021}T. Morita, Extracting Classical Lyapunov Exponent from One-Dimensional Quantum Mechanics, arXiv:2105.09603 [hep-th].

\bibitem{Banados1992}M. Banados, C. Teitelboim and J. Zanelli, The Black hole in three-dimensional space-time, Phys. Rev. Lett. 69, 1849 (1992).

\bibitem{Frassino2015}A.M. Frassino, R.B. Mann and J.R. Mureika, Lower-dimensional black hole chemistry, Phys. Rev. D 92, 124069 (2015).


\end{thebibliography}
\end{document}